\documentclass[twocolumn,secnumarabic,amssymb, nobibnotes, aps, pre, 10pt]{revtex4-2}

\usepackage{graphicx}
\usepackage{tikz}
\usepackage{enumitem}
\usepackage{amsmath}
\usepackage{extpfeil}
\usepackage{tikz}
\usetikzlibrary{shapes.geometric, arrows}

\tikzstyle{arrow} = [thick,->,>=stealth]

\usepackage{tikz}
\usetikzlibrary{shapes.geometric, arrows}

\tikzstyle{startstop} = [ellipse, minimum width=3cm, minimum height=1cm,text centered, draw=black]
\tikzstyle{process} = [rectangle, minimum width=3cm, minimum height=1cm, text centered, draw=black]
\tikzstyle{decision} = [diamond, minimum width=3cm, minimum height=1cm, text centered, draw=black]
\tikzstyle{arrow} = [thick,->,>=stealth]

\tikzstyle{startstop} = [rectangle, rounded corners, minimum width=3cm, minimum height=1cm,text centered, draw=black, fill=red!30]
\tikzstyle{process} = [rectangle, minimum width=3cm, minimum height=1cm, text centered, draw=black, fill=blue!20]
\tikzstyle{decision} = [diamond, minimum width=3cm, minimum height=1cm, text centered, draw=black, fill=green!20]
\tikzstyle{arrow} = [thick,->,>=stealth]

\begin{document}

\title{Minimum nonlinearity for pattern-forming Turing instability in a mathematical autocatalytic model}%

\author{Javier L\'opez-Pedrares}
\email{javier.lopez.pedrares@uvigo.gal}
\affiliation{Department of Mathematics, Universidade de Vigo, 36310 Vigo, Spain}
\affiliation{Group of Nonlinear Physics, Universidade de Santiago de Compostela, 15782 Santiago de Compostela, Spain}

\author{Marcos Su\'arez-V\'azquez}
\email{marcossuarez.vazquez@usc.es}
\affiliation{Galician Center for Mathematical Research and Technology (CITMAga), 15782 Santiago de Compostela, Spain}
\affiliation{Group of Nonlinear Physics, Universidade de Santiago de Compostela, 15782 Santiago de Compostela, Spain}

\author{Juan P\'erez-Mercader}
\email{jperezmercader@fas.harvard.edu}
\affiliation{Department of Earth and Planetary Sciences, Harvard University, Cambridge, MA 02138, USA}
\affiliation{Santa Fe Institute, Santa Fe, NM 87501, USA}

\author{Alberto P. Mu\~nuzuri}
\email{alberto.perez.munuzuri@usc.es} 
\affiliation{Galician Center for Mathematical Research and Technology (CITMAga), 15782 Santiago de Compostela, Spain}
\affiliation{Group of Nonlinear Physics, Universidade de Santiago de Compostela, 15782 Santiago de Compostela, Spain}
\affiliation{Department of Earth and Planetary Sciences, Harvard University, Cambridge, MA 02138, USA}


\begin{abstract}
	
Pattern formation is ubiquitous in nature and the mechanism widely-accepted to underlay them is based on the Turing instability, predicted by Alan Turing decades ago. This is a non-trivial mechanism that involves nonlinear interaction terms between the different species involved and transport mechanisms. We present here an analytical and numerical description aiming to explore the mathematical constraints that a model with a single autocatalytic term should comply in order to exhibit a Turing instability. The main conclusion limits the existence of this instability to nonlinearity degrees larger or equal to three. Additionally, the inclusion of cross-diffusion reduces the required nonlinearity degree by one, making instability possible at a lower order of nonlinearity.

\end{abstract}

\maketitle
\section{Introduction}
\label{sec:intro}

Since the seminal work proposed by Alan Turing \cite{turing52the}, it has become apparent that diffusion mechanisms play a crucial role in understanding the emergence of spontaneous pattern formation in active systems, a concept further refined by the activator-inhibitor principles of Gierer and Meinhardt \cite{gierer1972theory}. These pioneering insights demonstrated that reaction-diffusion coupling with the appropriate nonlinearity could give rise to self-organized structures through a symmetry-breaking process now widely known as the Turing instability \cite{kondo2010reaction}. Prigogine and Lefevre work in 1968 \cite{prigogine1968} proposed a simple mathematical two-variable model based on the Belousov-Zhabotinsky chemical oscillatory reaction \cite{zaikin1970} that exhibited Turing patterns \cite{Lefever2018}. Its first experimental validation was found in chemical systems decades later \cite{castets1990experimental}. This discovery established the fundamental theoretical framework to explain how complex patterns evolve in various natural systems \cite{murray2003mathematical}, ranging from nonlinear chemical reactions to morphogenesis in biological development \cite{koch1994biological}.

The experimental verification of this theoretical finding \cite{castets1990experimental} and posterior works \cite{de1991turing,dulos1996chemical,rudovics1999experimental} demonstrated the relevance of this phenomenon and the crucial role played in pattern formation and morphogenesis \cite{murray2007mathematical}. Indeed, the ability of such mechanisms to generate spatially ordered structures from initially homogeneous conditions had a profound impact on the understanding of developmental biology, tissue organization, and even ecological and chemical processes \cite{lengyel1992chemical}.

The so-called Turing mechanism is a clear non-trivial nonlinear phenomenon that can only be explained due to the high nonlinearity of the dynamical equations underlying the problems described. The interplay between local reaction kinetics and diffusion-driven instability leads to the spontaneous emergence of stable or oscillatory patterns, which would otherwise not be possible in the absence of these interactions.

In this work, we aim to understand the mathematical properties that the differential equations should exhibit in order to present a Turing instability. We focus on a particular set of partial differential equations that describe an autocatalytic reaction \cite{murray2007mathematical}. Autocatalytic processes are of critical importance in many processes in nature and, in particular, in living systems as they describe the way in which a living system produces more of its parts. These autocatalytic processes have been observed to play an important role in disciplines as diverse as chemistry, sociology, and economics \cite{horvath1999control,munuzuri1999control,mimar2019turing}.

The paper is organized as follows. First, the main mathematical results describing a Turing instability in a generic system, as well as the equations for a generic autocatalytic process in a reaction-diffusion context, are reviewed. In the next section, different nonlinearities are studied and the possibility of Turing instability is analyzed for each. Additionally, the impact of cross-diffusion terms in the equations is studied. Finally, the conclusions and a short discussion are presented.


\section{Turing instability}
\label{sec:turing_inst}

\subsection{Mathematical formulation}
\label{sec:turing_formulation}

Let us consider a generic reaction-diffusion system such as the one in the following equations,

\begin{equation}
		\frac{\partial u}{\partial t} = f(u,v) + D_u \nabla^2 u,
\end{equation}
\begin{equation}
	\frac{\partial v}{\partial t} = g(u,v) + D_v \nabla^2 v, 
\end{equation}

\noindent
where $f(u,v)$ and $g(u,v)$ account for the nonlinear interactions between the two species, $u$ and $v$, that fully describe the problem and that we suppose to be space-dependent variables. $D_u$ and $D_v$ are the diffusion coefficients responsible for the interactions between the different species spatially distributed describing the random interactions between both species. We consider them to be constant.

The Turing instability, as described in \cite{turing52the}, establishes that the system may have a fixed stable point in the absence of diffusion that becomes unstable for some range of values of the diffusion coefficients. When these two conditions are fulfilled, small inhomogeneous perturbations containing a specific wavelength and spatially distributed in an extended system, are amplified due to the diffusion effect and the system evolves into a spatially extended non-homogeneous structure with a well-defined wavelength that it is determined by the nonlinearity of the equations. These spatially-extended structures are called Turing patterns.

These two conditions can be expressed mathematically as follows. The first condition says that the system without the diffusion terms has a stable fixed point $(u_0, v_0)$. In terms of linear stability analysis, this means that the real-part of the two eigenvalues of the Jacobian matrix are negative \cite{strogatz2018nonlinear}. The Jacobian matrix, $A$, is given by,

\begin{equation}
    \begin{array}{c}
        \displaystyle
        A =\begin{pmatrix} 
            \frac{\partial}{\partial u} f(u,v) & \frac{\partial}{\partial v} f(u,v) \\  
            \frac{\partial}{\partial u} g(u,v) & \frac{\partial}{\partial v} g(u,v) 
        \end{pmatrix}= \\[10pt]
        \displaystyle
        \equiv 
        \begin{pmatrix} 
            f_u(u,v) & f_v(u,v) \\ 
            g_u(u,v) & g_v(u,v) 
        \end{pmatrix}.
    \end{array}
    \label{reac_diff}
\end{equation}

If $(u_0, v_0)$ is to be a stable fixed point, then the two following conditions must be fulfilled,

\begin{itemize}
	\item[-] \textit{\textbf{Condition 1}}: 
\end{itemize}

tr$(A) = f_u(u_0,v_0) + g_v(u_0,v_0) < 0$,
\begin{itemize}
	\item[-] \textit{\textbf{Condition 2}}: 
\end{itemize}

 $|A|= f_u(u_0,v_0)  g_v(u_0,v_0) - f_v(u_0,v_0)  g_u(u_0,v_0) > 0$.
\vspace{0.5cm}

In order to exhibit a Turing instability, another condition must be fulfilled when the diffusion is considered. The fixed point $(u_0,v_0)$ becomes unstable for some values of the diffusion coefficients for at least one wavelength on the initial extended perturbation. Following \cite{kapral2012chemical} this is translated in the following mathematical condition, 

\begin{itemize}
	\item[-] \textit{\textbf{Condition 3}}: 
\end{itemize}

$\Delta (m^2) = (f_u(u_0,v_0) - m^2D_u)(g_v(u_0,v_0) - m^2 D_v) < 0$,
\vspace{0.1cm}

\noindent
where $m$ is any arbitrary wavelength being part of the initial perturbation of the system. This condition is a convex quadratic function in $m^2$ (note that $D_v \; D_u>0$). $Condition \; 3$ implies, thus, that in order to exhibit a Turing instability, there must exist at least one wavelength, $m$, such that $\Delta (m^2)<0$ or, in other words, that the minimum of this quadratic expression is negative.

\subsection{General autocatalytic process}


Turing instability is a basic mechanism underlying pattern formation in living systems, which are primarily composed of individual subunits known as cells. From a fundamental perspective, cellular dynamics can be described by autocatalytic equations \cite{murray1993cell}, with a minimum of two variables, that schematically capture the ability of cells to synthesize their own components from available nutrients \cite{munuzuri2022unified}.



To mathematically capture these dynamics, it is easy to construct a model that incorporates two essential substances: the internal component  $V$, representing the components of the living system, and the external nutrient  $U$.  Within this framework, $V$ undergoes reactions with $U$ to produce new  $V$ components, allowing the system growth, while a decay reaction eventually transforms  $V$ into an inert waste substance  $C$.



	
	
	

A straightforward kinetic model that meets the above constraints in the most general way is,




\begin{eqnarray}
    \text{Reservoir}\; \xlongrightarrow{\quad F \quad}\; &U&, \nonumber \\
    \alpha U + \beta V\; \xlongrightarrow{\quad \lambda \quad}\; (\alpha - 1)U + (\beta + 1)&V&, \nonumber \\
    V\; \xlongrightarrow{\quad k \quad}\; &C&~.
\end{eqnarray}

This set of reactions can be translated, using the Mass Action Law \cite{moore1972physical,atkins1986physical}, to the following set of partial differential equations (PDEs) that have an $\alpha+\beta$ non-linearity,

\begin{eqnarray}
		\frac{\partial u}{\partial t} &=& f(u,v) + D_u \nabla^2 u = \nonumber \\ 
		&=& -\lambda u^{\alpha}v^{\beta} + F(1-u) + D_u \nabla^2 u \label{eqn:u}, \\
		\frac{\partial v}{\partial t} &=& g(u,v) + D_v \nabla^2 v = \nonumber \\
		&=& \lambda u^{\alpha}v^{\beta} - (F+k)v + D_v \nabla^2 v \label{eqn:v}~.
		\label{general_system}
\end{eqnarray}

Note that the first equation describes the feeding process where species $u$ is introduced into the system from the reservoir at rate \( F \). However, the partial differential equations in Equations \eqref{eqn:u} and \eqref{eqn:v} currently do not account for the outflow of species $u$ and $v$ at this rate. In the reaction mechanism, both $u$ and $v$ are not only produced and transformed but also removed from the system at rate \( F \). To correctly reflect this, the PDEs should include additional terms representing the outflow of $u$ and $v$, such as \( -F u \) and \( -F v \), to capture the loss of these species from the system.

As $u$ and $v$ species are spatially distributed, we suppose that the transport mechanism is purely diffusive, although more complex transport mechanism could be considered.


\section{Results}
\label{sec:results}

In this section, we start from the general autocatalytic reaction-diffusion set of equations introduced above and analyze its viability to exhibit a Turing instability as described in the previous section. For that, we will analyze the different configurations of the nonlinear terms and their influence in the existence of the instability.

Without any loss of generality, we will set $\alpha=1$ in the following. The parameter $\alpha$ describes the number of molecules of type ``food'' needed to trigger the autocatalysis. Note that for $\alpha>1$ we can always consider that the reaction takes place when $\beta$ number of $v$ particles interact with a single particle composed by $\alpha$ $u$-particles, thus, it becomes equivalent to the case $\alpha=1$. From now on we will focus on nonlinear terms like $u v^{\beta}$.







The resulting system may present multiple fixed points (named ($u_0, v_0$) generically) depending on the values of the parameters. It is easy to see that there is one simple fixed point that it is always present independently of the nonlinearity considered or the parameter values:  $(u_0, v_0) = (1, 0)$. The stability of a fixed point (without diffusion) is calculated through the Jacobian matrix,

\begin{equation}
    \begin{array}{c}
        \displaystyle
     A \equiv 
        \begin{pmatrix} 
            f_u(u_0,v_0) & f_v(u_0,v_0) \\ 
            g_u(u_0,v_0) & g_v(u_0,v_0) 
        \end{pmatrix} =\\[10pt]
        \displaystyle
        = \begin{pmatrix} 
            -\lambda v_0^{\beta} - F & -\lambda \beta u_0 v_0^{\beta-1} \\ 
            \lambda v_0^{\beta} & \lambda\beta u_0 v_0^{\beta - 1} - (F+k) 
        \end{pmatrix}.
    \end{array}
    \label{eq:jac_matrix}
\end{equation}

Note that the properties of the above matrix completely depend on the actual value of the nonlinear exponent $\beta$ ($\beta \in \mathbb{N}$). Thus, we will study three different subcases in the coming analysis: $\beta = 0, \ \beta = 1$ and $\beta \geq 2$.


\subsection{Case $u^{\alpha}$ ($\beta = 0$)}

This is a non-relevant case as it may imply the spontaneous production of constituents from ``food'' without autocatalysis but, for completeness it is analyzed here.





The first Turing condition states that the fixed point $(u_0, v_0)$ should be stable in absence of diffusion terms. The stability of a fixed point can be calculated through the Jacobian matrix, that in this case becomes,

\begin{equation}  
    \begin{array}{c}
      A \equiv    \displaystyle
        \begin{pmatrix} 
            f_u(u_0,v_0) & f_v(u_0,v_0) \\  
            g_u(u_0,v_0) & g_v(u_0,v_0) 
        \end{pmatrix} =\\[10pt]
        \displaystyle
        = 
        \begin{pmatrix} 
            -2\lambda \alpha u_0^{\alpha - 1} - F & 0 \\  
            2\lambda \alpha u_0^{\alpha - 1} & -(F + k) 
        \end{pmatrix}.
    \end{array}
\end{equation}

The following two conditions must be fulfilled,
\begin{itemize}
	\item[-] \textit{\textbf{Condition 1}}:
\end{itemize}    

tr$(A) = f_u + g_v = -2\lambda\alpha u_0^{\alpha-1} - F - (F+k) < 0$,

\begin{itemize} 
	\item[-] \textit{\textbf{Condition 2}}: 
\end{itemize}

$|A| = f_ug_v - f_vg_u = (2\lambda\alpha u_0^{\alpha-1} + F)(F+k) > 0$.
\vspace{0.5cm}

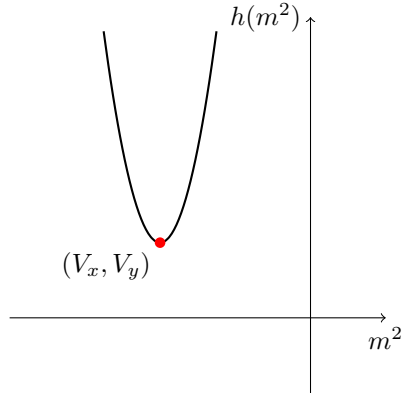
\begin{figure}[t!]
	\centering
	\begin{tikzpicture}
		\draw[->] (-4,0) -- (1,0) node[anchor=north] {$m^2$};
		\draw[->] (0,-1) -- (0,4) node[anchor=east] {$h(m^2)$};
		
		\draw[thick,domain=-2.75:-1.25,smooth,variable=\x] plot ({\x},{5*(\x+2)*(\x+2) + 1});
		
		\fill[red] (-2,1) circle (2pt);
		\node[anchor=north east] at (-2,1) {$(V_x, V_y)$};
	\end{tikzpicture}
	\caption{Sketch of the quadratic function obtained from \textbf{Condition 3}.}
	\label{fig:parabola}
\end{figure}

Both conditions are always fulfilled as all the parameters are defined positive. The second Turing condition states that the fixed point $(u_0, v_0)$ should turn unstable when the diffusion terms appear. This is translated for the present case as described in Section~\ref{sec:turing_formulation} in the following inequality,


\begin{itemize}
	\item[-] \textit{\textbf{Condition 3}}:
\end{itemize}
\begin{align*}
    \Delta (m^2) &= (f_u - m^2 D_u)(g_v - m^2 D_v) \\
    &= f_u g_v - m^2 (D_v f_u + D_u g_v) + D_v D_u m^4 < 0,
\end{align*}
\vspace{0.1cm}

\noindent
which corresponds to a convex quadratic function in $m^2$, because $D_vD_u>0$. This means the function has a global minimum at its vertex, whose $y$ coordinate ($V_y$) should be negative to fulfill the condition. We can calculate the vertex coordinates as follows, 

\begin{eqnarray}
		V_x &=& - \frac{-(D_vf_u + D_ug_v)}{2D_uD_v} < 0, \nonumber \\
		V_y &=& f_ug_v - m^2(D_vf_u + D_ug_v) + \nonumber 
		+ D_vD_um^4 > 0,
\end{eqnarray}

\noindent
because $D_vf_u + D_ug_v$ is always negative. This means that the parabola is always positive, as in Figure \ref{fig:parabola}, so \textbf{Condition 3} is never fulfilled independently on the fixed point considered and, thus, \textbf{Condition 3} is not fulfilled and the Turing instability is not present in this system.


\subsection{Case $u v$ ($\beta = 1$)}

We analyze in this case the simplest autocatalytic process with a quadratic nonlinearity.
The fixed points of the system without diffusion can be easily calculated by equating the reactive part of equation \eqref{general_system} to zero,





\begin{eqnarray}
		0 &=& -\lambda uv + F(1 - u), \nonumber \\
		0 &=& \lambda uv - (F + k)v.
\end{eqnarray}

The solutions for this set of coupled equations are,

\begin{enumerate}[label=(\roman*)]
	\item $(u_0, v_0) = (1,0)$.
	\item $(u_0,v_0) = \left( \dfrac{F + k}{\lambda}, \dfrac{F}{\lambda}\dfrac{\lambda - F - k}{F + k}\right)$.
\end{enumerate}

The Jacobian matrix for the system is,
\begin{equation}
     \begin{array}{c}
        A \equiv \displaystyle
        \begin{pmatrix} 
            f_u(u_0,v_0) & f_v(u_0,v_0) \\  
            g_u(u_0,v_0) & g_v(u_0,v_0) 
        \end{pmatrix}= \\[10pt]
        \displaystyle
        =\begin{pmatrix} 
            -2\lambda \alpha u_0^{\alpha - 1} - F & 0 \\  
            2\lambda \alpha u_0^{\alpha - 1} & -(F + k) 
        \end{pmatrix}.
    \end{array}
\end{equation}

Let's check the Turing conditions for the different equilibrium points,

\begin{enumerate}[label=(\roman*)] 
	\item $(u_0,v_0) = (1, 0)$,
	\begin{itemize}
		\item[-] \textit{\textbf{Condition 1}}:
    \end{itemize}
    tr$(A)=f_u + g_v = - F + \lambda - F - k < 0 \Rightarrow \lambda < 2F + k$.
    \begin{itemize}
		\item[-] \textit{\textbf{Condition 2}}:
    \end{itemize}
    $|A|=f_ug_v - f_vg_u = f_ug_v > 0 \Rightarrow \lambda < F + k$, \\ which is more restrictive than \textit{\textbf{Condition 1}}, as the parameters are defined positive.
    \begin{itemize}
		\item[-] \textit{\textbf{Condition 3}}:
    \end{itemize}
    $-\dfrac{\left(D_vf_u + D_ug_v \right)^2}{4D_uD_v} + f_ug_v - f_vg_u = \\ = -\dfrac{\left(-D_vF - D_u(-\lambda + F + k)\right)^2}{4D_uD_v} - F(\lambda - F - k) < 0  \Rightarrow \lambda > F + k $, \\ which is the opposite to \textbf{Condition 2}, thus, Turing instability is impossible.

	\item  $(u_0,v_0) = \left( \dfrac{F + k}{\lambda}, \dfrac{F}{\lambda}\dfrac{\lambda - F - k}{F + k}\right)$,
	\begin{itemize} 
		\item[-] \textit{\textbf{Condition 1}}:
    \end{itemize}
    tr$(A)=f_u + g_v = -\dfrac{F\lambda}{F + k} < 0$.
    \begin{itemize}
		\item[-] \textit{\textbf{Condition 2}}:
    \end{itemize}
    $|A| = f_ug_v - f_vg_u = -f_vg_u > 0 \Rightarrow f_vg_u < 0 \Rightarrow \\ \Rightarrow -(F+k)F\dfrac{\lambda-F-k}{F+k} \Rightarrow \lambda > F + k$, \\ which is the opposite condition to the one for the other fixed point.
    \begin{itemize}
		\item[-] \textit{\textbf{Condition 3}}:
    \end{itemize}
    $-\dfrac{\left(D_vf_u + D_ug_v \right)^2}{4D_uD_v} + f_ug_v - f_vg_u = \\ = -\dfrac{\left(-D_v\frac{F\lambda}{F+k}\right)^2}{4D_uD_v} - F(\lambda - F - k) < 0 \Rightarrow \\ \Rightarrow \lambda < F + k $, \\ which is the opposite to Condition 2 and, thus, Turing instability is not possible either for this case.
\end{enumerate}

For both fixed points, conditions 2 and 3 are contradictory, so there is no Turing instability for this type of nonlinear system. Also, for both fixed points, exhaustive 2D numerical integration of the reaction-diffusion equations was done and Turing instability was never observed. We conclude that Turing structures cannot appear in this system.

\begin{figure*}[ht!]
	\centering
	\includegraphics[width=1.0\textwidth]{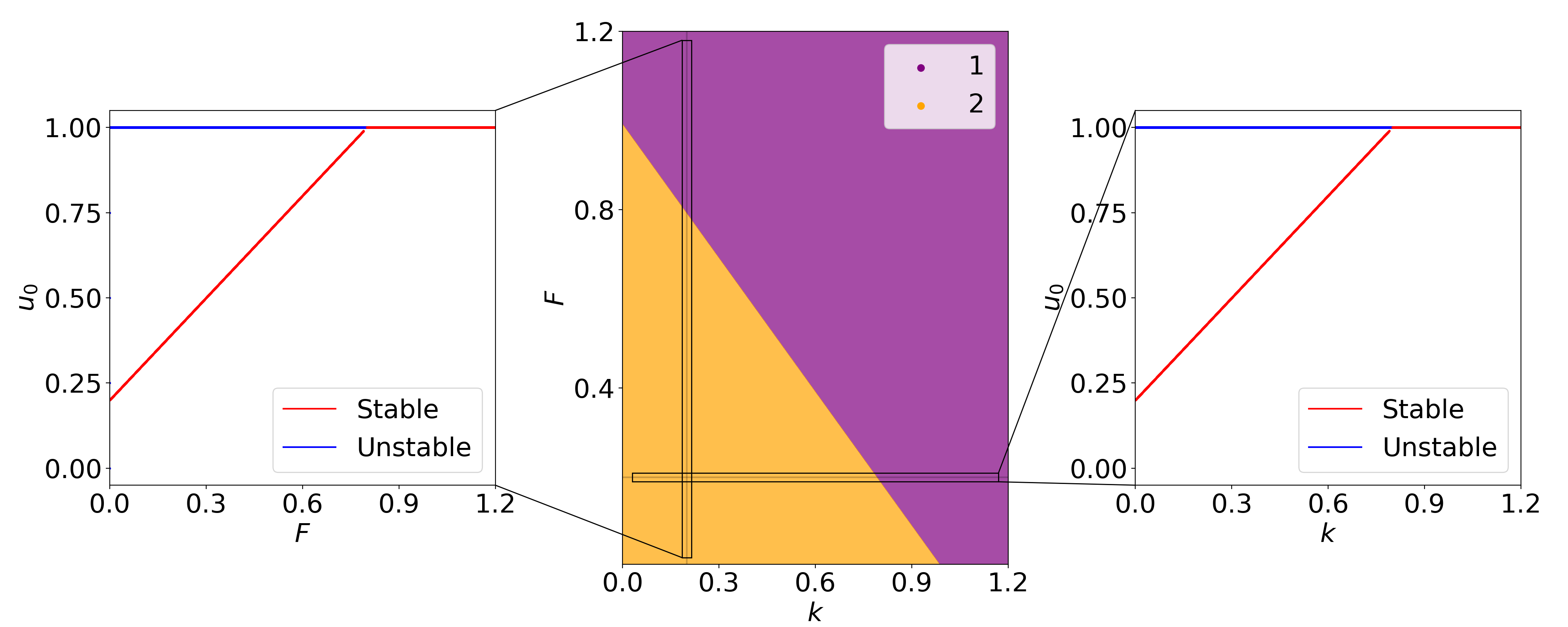}
	\caption{Bifurcation diagram for $\beta = 1$. The central plot shows the number of real solutions of the system for a range of $F$ and $k$ values ($\lambda = 1$). The side plots represent the stability of the fixed points for the highlighted regions varying $F$ and $k$ respectively.}
	\label{bifurcation_beta1}
\end{figure*}

Figure \ref{bifurcation_beta1} shows a bifurcation diagram for the case $\beta = 1$. Notice that for higher $F$ and $k$ values the only fixed point is $(u_0, v_0) = (1, 0)$, so no Turing behavior is observed. For lower values in the orange region, where $\lambda > F + k$, the stability conditions change and two fixed points appear in the system, but the third Turing condition is still not satisfied, so no Turing instability is observed.

\subsection{Case $u v^{\beta}$ \textnormal{($\beta \geq 2$)}}

This case with $\alpha = 1$ and $\beta = 2$ resembles the Gray-Scott model \cite{doelman1997pattern,mcgough2004pattern} and the Brusselator model introduced in \cite{prigogine1968}. Autocatalytic processes with this type of nonlinearity always have a fixed point at $(u_0, v_0) = (1,0)$. This point does not fulfill the Turing conditions as we can easily derive from its Jacobian matrix,


\begin{equation}
	A \left((1, 0)\right) = \begin{pmatrix} -F & 0 \\ 0 & -(F + k) \end{pmatrix}.
	\label{jaco}
\end{equation}

As stated before, the Turing instability requires the following inequalities,

\begin{itemize}
	\item[-] \textit{\textbf{Condition 1}}:
\end{itemize}    
    tr$(A)=f_u + g_v = - F - (F + k) < 0$.
\begin{itemize}
	\item[-] \textit{\textbf{Condition 2}}:
\end{itemize}    
    $|A|=f_ug_v - f_vg_u = F(F + k) > 0$ .
	
\begin{itemize}
	\item[-] \textit{\textbf{Condition 3}}: 
\end{itemize}
$\Delta (m^2) = (f_u - m^2D_u)(g_v - m^2 D_v) < 0$, 

\noindent
which is not verified.

Conditions 1 and 2 are verified just by looking at the values of the Jacobian matrix in Equation \eqref{jaco}. Condition 3, on the other hand, is never fulfilled, as $f_u$ and $g_v$ are always negative. This means that the parabola in $m^2$ is always positive independently on the wavelength, $m$, considered. Thus, this fixed point cannot give rise to a Turing structure.

In the following, we will analyze the remaining fixed points for the different nonlinearity degrees. As $\beta$ grows larger, the analytical calculation of the additional fixed points becomes more non-trivial. The Supplementary Information shows the analytical calculations of the fixed points for the cases with $\beta = 1, 2$.

\begin{figure*}[ht!]
	\centering 
    \includegraphics[width=\textwidth, angle=0]{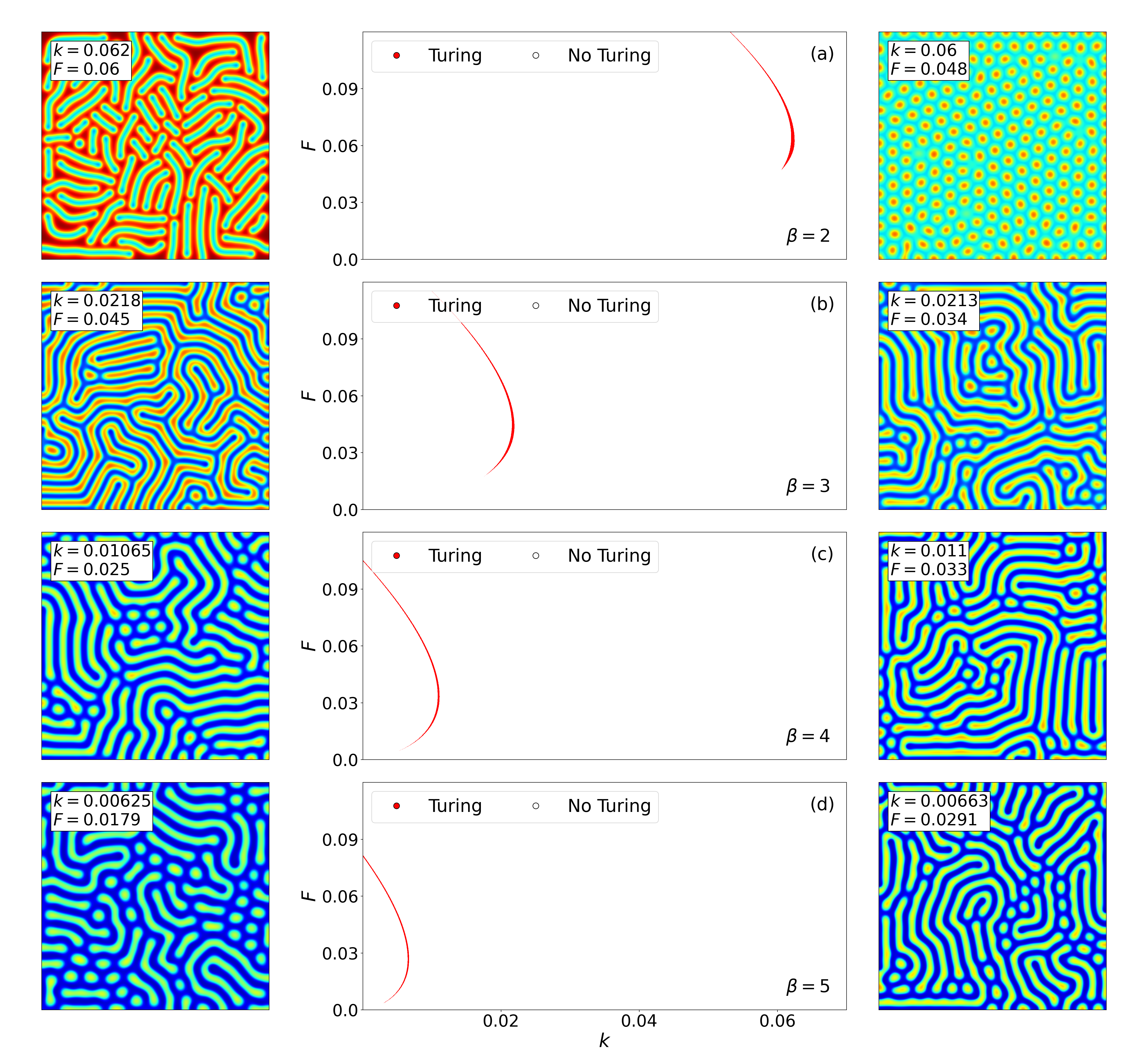}	
	\caption{Numerical calculation of the Turing conditions for (a) $\beta = 2$, (b) $\beta = 3$, (c) $\beta = 4$, (d) $\beta = 5$. Panels in the middle present the $k$-$F$ phase diagrams marking the parameter range where Turing conditions are fulfilled. The lateral panels present direct numerical simulations of the equations proving the existence  of Turing patterns inside the predicted regions. We used the three-level fully-explicit Dufort-Frankel integration scheme on a $500 \times 500$ spatial lattice and we waited $10^6$ time iterations till the stationary solution was obtained ($\Delta t = 0.1$, $\Delta x = 1$, $\lambda = 1$, $D_u = 1.0$ and $D_v = 0.5$).} 
	\label{CompleteSims}
\end{figure*}

In Figure \ref{CompleteSims} we present a numerical demonstration that Turing conditions are fulfilled for values of $\beta$ from $\beta=2$ up to $\beta=5$. The central panels present the $F$-$k$ phase portraits marking in red the parameter ranges where the Turing conditions are fulfilled. Here, for each nonlinearity, the fixed points are numerically calculated (see details in the Supplementary Material) and the Turing conditions tested for each value in the $F$-$k$ phase diagram. Lateral columns show numerical evidences of Turing structures after direct numerical  integration of the set of equations \eqref{general_system} for some specific values of $F$ and $k$ ($\lambda = 1$, $D_u = 1.0$ and $D_v = 0.5$). In all the cases analyzed, the results in the phase portraits were numerically verified. Note that only the spacial distribution of the $u$-values are plotted in this figure (details on the $v$-variable as well as some additional examples can be observed in the Supplementary Information). 
The results with $\beta = 2$ are coincident with those in \cite{munuzuri2022unified} as well as in previous papers \cite{hochberg2003large,hochberg2005spatiotemporal}, but larger values of $\beta$ were still unexplored in the literature.

The same calculations can be done for higher order nonlinearity in Equations \eqref{general_system} by just considering a larger value for $\beta$ and redoing the numerical approach presented here. In all cases the Turing instability are fulfilled and Turing structures can be observed.

\subsection{Cross-diffusion effect on nonlinearity degree}
\label{sec:cross_diff}

In this section we analyze the effect of cross-diffusion  \cite{mimar2019turing,vanag2009cross, wang2011complex, xue2012pattern, ruiz2013mathematical, vidal2017social} on the effective nonlinearity of the system considered. Cross-diffusion naturally arises when the diffusion coefficient depends on the value of the system variables \cite{vanag2009cross}. For that, let us suppose a generic two-variable nonlinear reaction-diffusion system given by,

\begin{equation}
\begin{split}
	\frac{\partial u}{\partial t} &= f(u, v) + \nabla^2 (D_u u) \\
	\frac{\partial v}{\partial t} &= g(u, v) + \nabla^2 (D_v v).
\end{split}
\end{equation}

In general, the diffusion coefficients may also depend on the actual concentration of the different species involved $(u, v)$ \cite{vanag2009cross},

\begin{equation}   
D_i = D_i(u, v) \simeq D_i^0 + D_{iu}^0 u + D_{iv}^0 v,
\end{equation}

\noindent
where $D_i^0, D_{iu}^0, D_{iv}^0 \in \mathbb{R}$ are constants, independent of $u$ and $v$, and with $i=u,v$. Thus, the set of reaction-diffusion equations becomes,

\begin{eqnarray}
	\frac{\partial u}{\partial t} &=& f(u, v) + \nabla^2 \left( \left( D_u^0 + D_{uu}^0 u + D_{uv}^0 v \right) u \right), \nonumber \\
	\frac{\partial v}{\partial t} &=& g(u, v) + \nabla^2 \left( \left( D_v^0 + D_{vu}^0 u + D_{vv}^0 v \right) v \right),
\end{eqnarray}

\noindent
where $D_{ii}^0$ are known as self-diffusion coefficients and they reinforce normal diffusion. $D_{ij}^0$ with $i \neq j$ are named cross-diffusion coefficients and express the influence on the diffusion of a given species by the existence of gradients of the other species involved.

These dynamical equations are used to obtain the values of $(u, v)$ at a given time as a function of the previous instant of time following an integration scheme such as,

\begin{eqnarray}
	u(t + \Delta t, &x&, y) \simeq u(t, x, y) + \Delta t \left\{ f(u, v) + \right. \nonumber \\
	&+& \left. \nabla^2 \left( \left( D_u^0 + D_{uu}^0 u + D_{uv}^0 v \right) u \right) \right\}, \nonumber \\
	v(t + \Delta t, &x&, y) \simeq v(t, x, y) + \Delta t \left\{ g(u, v) + \right. \nonumber \\
	&+& \left. \nabla^2 \left( \left( D_v^0 + D_{vu}^0 u + D_{vv}^0 v \right) v \right) \right\}~.
\end{eqnarray}

From these equations, one can observe that the values of $(u, v)$ are proportional to the nonlinear functions $f$ and $g$. Thus, the cross-diffusion terms where $u$ and $v$ explicitly appear also include this nonlinearity, 

\begin{eqnarray}
	D_{uv}^0 \, v \, u \propto g(u, v) \, u, \nonumber \\
	D_{vu}^0 \, u \, v \propto f(u, v) \, v~.
\end{eqnarray}

In both cases, the effective non-linearity is increased by one degree in both equations. The meaning of this justification in this context is that a second order nonlinearity---i.e., a system with linear kinetics, as seen in \cite{vanag2009cross}---in an autocatalytic reaction can exhibit Turing instability if cross-diffusion is present.


\section{Summary and conclusions}
\label{sec:conclusions}

In this work, we analyzed the minimal mathematical constraints that an autocatalytic process must confer in order to exhibit a Turing instability. This instability is believed to be behind many pattern-forming mechanisms in nature and our results confirm that a cubic nonlinearity is the minimum compatible with the emergence of Turing patterns in such systems.


For systems with lower-order nonlinearity, $\beta< 2$, we found that all Turing conditions could not be simultaneously fulfilled, preventing the formation of Turing patterns within the system. On the other hand, for systems with higher-order nonlinearity, $\beta\geq 2$, specific parameter behaviors were identified where Turing conditions are met. Numerical simulations validated the theoretical predictions, with distinct Turing patterns emerging in these parameter spaces. These results confirm that increasing nonlinearity significantly enhances the system's ability to exhibit spatially extended instabilities.

We also analyzed the role played by cross-diffusion in order to increase the effective nonlinearity of the system and its role in facilitating the possibility to exhibit Turing patterns. We demonstrate that higher order diffusion coefficients increase the system complexity and, thus, favor the appearance of Turing instability. 

Turing instability, thus, results as a conclusion of a minimum complexity that may come directly through the nonlinear interactions between the different species involved but also via the more complex diffusion coefficients (cross-diffusion).

We believe that our results developed in the context of autocatalytic processes could also apply to different models although additional research should be done to confirm it.

These findings have significant implications for understanding the mechanisms underlying pattern formation in biophysical systems. The results emphasize the importance of nonlinear dynamics in driving complex spatial structures and offer a predictive framework for exploring Turing instabilities in autocatalytic processes.

\section*{Declaration of competing interest}
The authors declare that they have no known competing financial interests or personal relationships that could have appeared to influence the work reported in this paper.

\section*{Data availability}
No data was used for the research described in the article. 

\section*{Acknowledgments}
We gratefully acknowledge financial support by the Spanish Ministerio de Ciencia e Innovación and European Regional Development Fund under contract PID2022-138322OB-I00 AEI/FEDER, UE, and by Xunta de Galicia under Research Grant No. 2025-PG025. All these programs are co-funded by FEDER (UE). MSV is thankful for the support by the Doutoramento Industrial grant from Xunta de Galicia (IN606D). JPM thanks Repsol, SA and Harvard’s Origins of Life Initiative for their support.

\bibliography{sample_turing.bib}

\end{document}